# Statistics of the first passage time of Brownian motion conditioned by maximum value or area

Michael J Kearney

*Faculty of Engineering and Physical Sciences,*

*University of Surrey, Guildford, Surrey, GU2 7XH, UK*

Satya N Majumdar

*Laboratoire de Physique Théorique et Modèles Statistiques,*

*Université Paris-Sud. Bât. 100, 91405, Orsay Cedex, France*

September 2014

Abstract

We derive the moments of the first passage time for Brownian motion conditioned by either the maximum value or the area swept out by the motion. These quantities are the natural counterparts to the moments of the maximum value and area of Brownian excursions of fixed duration, which we also derive for completeness within the same mathematical framework. Various applications are indicated.

## 1. Introduction

A Brownian excursion $B_{ex}(\tau)$ is a realisation of the Wiener process which is suitably conditioned so that $B_{ex}(0) = B_{ex}(t) = 0$ with $B_{ex}(\tau) > 0$ for $0 < \tau < t$, where $t$ is fixed. The moments of the maximum value $M_{ex} = \max_{0 \leq \tau \leq t} B_{ex}(\tau)$ of such excursions are well-known and follow quickly from the exact distribution given in [1-3], see also [4]. Similarly, the moments of the area functional $A_{ex} = \int_0^t B_{ex}(\tau) d\tau$ are also well-known [5-8]. The latter, in particular, find a wealth of applications in physics, mathematics and computer science, diverse examples of which include the statistics of the maximal relative height of fluctuating interfaces and the cost of constructing a table for data storage using linear probing with a random hashing algorithm (for an overview see [7, 8] and references therein). There is also a natural link in the limit to various discrete combinatorial problems occurring in graph theory that are related to Bernoulli processes (see [9] and references therein).

The nature of the Wiener process means there are certain mathematical subtleties with enforcing the positivity condition $B_{ex}(\tau) > 0$ for $0 < \tau < t$ given that $B_{ex}(0) = B_{ex}(t) = 0$. One way to deal with this is to consider a more general Brownian motion $X(\tau) = x + W(\tau)$, where $W(\tau)$ is the Wiener process with $W(0) \equiv 0$ and $x > 0$ is the initial condition. Let $T = \inf\{\tau : X(\tau) = 0\}$ denote the first passage time to the origin, $M = \max_{0 \leq \tau \leq T} X(\tau)$ the maximum value reached during the first passage, and $A = \int_0^T X(\tau) d\tau$ the first passage area functional. Then the conditional expectations $E_x(M^n|T=t)$ and $E_x(A^n|T=t)$, evaluated in the limit $x \to 0$, provide the corresponding moments for the maximum value and area of Brownian excursions.

This approach suggests a natural counterpart which forms the topic of this paper, namely the calculation of the conditional expectations $E_x(T^n|M=m)$ and $E_x(T^n|A=a)$ in the limit $x \to 0$. Unlike for excursions, whose duration is fixed, these quantities relate to Brownian motions $X(\tau)$ whose first passage time is variable but which are conditioned to have a given maximum value or area. In the former case, the result may be inferred from the remarkable relationships that exist between Brownian processes and the Riemann zeta function outlined in [10]. However, the derivation presented in Section 2, which we physically motivate in terms of a refinement of the classic escape problem for Brownian motion on a finite interval with two absorbing boundaries [11, 12], provides a different perspective and one which is much more transparent. In the latter case, the derivation is a more challenging task and less easy to motivate physically, although in Section 3 we point out a connection to the random acceleration process whose inherently non-Markovian characteristics account for the more involved nature of the calculation (for a wider discussion of first passage properties in non-Markovian systems see [13]). Additionally, in both cases we also show how the counterpart results for standard Brownian excursions can be derived within the same mathematical framework. This unified approach is not just of intrinsic interest, it is also relevant in the context of generic estimation problems, as discussed in Section 4 where we also offer some concluding remarks.

## 2. Motion conditioned by maximum value

The conditional expectation $E_x(T^n | M = m)$ can be derived from the joint probability density $P(m, t, x)$ associated with the distribution $\Pr(M \leq m, T \leq t, x)$. As a preliminary we introduce the Laplace transformed quantity,

$$Q(m, p, x) = \int_0^\infty e^{-pt} \Pr(M \leq m, T = t, x) \, dt \tag{1}$$

which, through the use of standard Feynman-Kac methods may be shown to satisfy the backward Fokker-Plank equation (see Appendix);

$$\left[ \frac{1}{2} \frac{\partial^2}{\partial x^2} - p \right] Q(m, p, x) = 0. \tag{2}$$

The relevant boundary conditions are $Q(m, p, x = 0) = 1$ and $Q(m, p, x = m) = 0$, whereupon the solution is,

$$Q(m, p, x) = \frac{\sinh(\sqrt{2p}(m - x))}{\sinh(\sqrt{2p}m)}. \tag{3}$$

The poles of $Q(m, p, x)$ are located on the negative real axis at $p = -\pi^2 k^2 / 2m^2$, where $k = 1, 2, 3...$, so (3) can be inverted to give,

$$\Pr(M \leq m, T = t, x) = \frac{\pi}{m^2} \sum_{k=1}^{\infty} (-1)^{k+1} k \sin\left(\frac{k\pi(m-x)}{m}\right) e^{-\frac{k^2\pi^2}{2m^2}t}$$

(4)

$$= \frac{\pi}{m^2} \sum_{k=1}^{\infty} k \sin\left(\frac{k\pi x}{m}\right) e^{-\frac{k^2\pi^2}{2m^2}t}.$$

The probability $\Pr(M \leq m, x) \equiv \int_0^\infty \Pr(M \leq m, T = t, x)\, dt$ that a given realisation of the first passage process does not exceed a maximum value $m$ can be easily obtained from (1) and (3);

$$\Pr(M \leq m, x) = \lim_{p \to 0} Q(m, p, x) = 1 - \frac{x}{m}. \tag{5}$$

This result is well-known, see e.g. [14]. It follows immediately that one can write down using (1), (3) and (5),

$$E_x(e^{-pT} | M \leq m) \equiv \frac{\int_0^\infty e^{-pt} \Pr(M \leq m, T = t, x)\, dt}{\Pr(M \leq m, x)}$$

(6)

$$= \left(\frac{m}{m-x}\right) \frac{\sinh(\sqrt{2p}(m-x))}{\sinh(\sqrt{2p}\, m)}.$$

From this one can derive $E_x(T | M \leq m)$, $E_x(T^2 | M \leq m)$ etc. by expanding both sides of (6) as a power series in $p$ and comparing the coefficients of $p$, $p^2$ etc. This yields for the first two conditional moments,

$$E_x(T|M \leq m) = \frac{x(2m-x)}{3};$$

$$E_x(T^2|M \leq m) = \frac{x(2m-x)}{45}(4m^2 + 6mx - 3x^2).$$
(7)

The first of these is simply the expected escape time for a particle undergoing Brownian motion which starts at position $x$ on the interval $[0,m]$ and exits the interval via the origin. The result is standard, but the derivation here is somewhat more straightforward than the conventional approach which relies on defining probability currents and so forth [11, 12].

A refinement of this classic escape problem may be introduced as follows. Noting that $P(m,t,x) \equiv \frac{\partial}{\partial m}\Pr(M \leq m, T = t, x)$, from (1) and (3) one can derive the Laplace transform with respect to $t$ of the probability density $P(m,t,x)$;

$$\int_0^\infty e^{-pt} P(m,t,x)\,dt = \frac{\partial Q}{\partial m} = \sqrt{2p}\,\frac{\sinh(\sqrt{2p}\,x)}{\sinh^2(\sqrt{2p}\,m)}.$$
(8)

If we let $P(m,x) \equiv \int_0^\infty P(m,t,x)\,dt$ denote the probability density of the maximum value, it follows from (8) that

$$P(m,x) = \lim_{p \to 0} \frac{\partial Q}{\partial m} = \frac{x}{m^2}.$$
(9)

Naturally this could also be derived directly from (5), as was done in [14]. Using (8) and (9) one can therefore write down,

$$E_x(e^{-pT}|M=m) \equiv \frac{\int_0^\infty e^{-pt} P(m,t,x)\,dt}{P(m,x)}$$

$$= \sqrt{2p}\,\frac{m^2}{x}\,\frac{\sinh(\sqrt{2p}\,x)}{\sinh^2(\sqrt{2p}\,m)}.$$

(10)

As above, by expanding both sides of (10) as a power series in $p$ and comparing coefficients one can evaluate $E_x(T^n|M=m)$ for $n=1,2,3,\ldots$ etc. For the first two conditional moments one has;

$$E_x(T|M=m) = \frac{2m^2 - x^2}{3};$$

$$E_x(T^2|M=m) = \frac{24m^4 - 20m^2 x^2 + 3x^4}{45}.$$

(11)

In contrast to (7), the first of these results is now the expected escape time for a particle undergoing Brownian motion which starts at position $x$ on the interval $[0,m]$ and exits the interval via the origin, but before doing so explores the whole of the interval. In figure 1 and figure 2 we compare (on a logarithmic scale) the expectation and standard deviation $\sigma_m(T) \equiv \sqrt{\mathrm{Var}(T)}_{M=m}$ based on (11), with initial condition $x=1$, against numerical averages derived from simulating many realisations of the process. The agreement is excellent. Such simulations are easy to perform by appealing to the fact that the limiting behaviour of a suitably scaled random walk converges in distribution to Brownian motion; thus a given realisation (say the $i$-th) is generated according to the iterative scheme $X_{n+1}^{(i)} = X_n^{(i)} \pm \sqrt{\Delta\tau}$, where $\Delta\tau$ is a small time increment and the sign of the increment is chosen randomly with equal probability at each step. The process evolves until step $N^{(i)}$ where $X_n^{(i)} < 0$ for the first time, whereupon $T^{(i)} = N^{(i)} \Delta \tau$ and $M^{(i)} = \max_{0 \leq n < N^{(i)}} \{X_n^{(i)}\}$. Anticipating what is to come in the next section, the area functional $A^{(i)} = \Sigma_{n=0}^{n=N^{(i)}} X_n^{(i)} \Delta \tau$.

When $x = m$, the results (7) and (11) coincide, as of course they must on physical grounds since they both describe a particle undergoing Brownian motion which starts at the extreme end of the interval $[0, m]$ and exits the origin after time $T$. By reflection symmetry, this is identical to the time it takes for a so-called Brownian meander to start from the origin and reach a threshold level $m$ [7, 9]. From (10) we have that,

$$E_{x=m}(e^{-pT} | M = m) = \frac{\sqrt{2pm}}{\sinh(\sqrt{2pm})} = 1 + \sum_{n=1}^{\infty} \frac{2^{n+1}(1 - 2^{2n-1})B_{2n} m^{2n}}{(2n)!} p^n \qquad (12)$$

where $B_n$ is the $n$-th Bernoulli number. By comparing the coefficients of $p$ it therefore follows that,

$$E_{x=m}(T^n | M = m) = \frac{4(2^{2n-1} - 1)n!}{2^n \pi^{2n}} \zeta(2n) m^{2n} \qquad (13)$$

where $\zeta(s) \equiv \Sigma_{k=1}^{\infty} k^{-s}$ is the Riemann zeta function. In writing (13), which gives the moments of the first passage time (or threshold hitting time) for a Brownian meander, we have used the well-known connection between $B_{2n}$ and $\zeta(2n)$ [15; Sect. 23.2.16].

An alternative perspective on this result, which provides a deep insight into the connection between Brownian processes and the Riemann zeta function, may be found in [10].

In the limit $x \to 0$, on the other hand, one has from (10) and (12),

$$E_0(e^{-pT}|M=m) = \frac{2pm^2}{\sinh^2(\sqrt{2pm})} = \left[E_{x=m}(e^{-pT}|M=m)\right]^2. \tag{14}$$

This has an obvious interpretation; the motion is now a double meander, where the two contributing meanders (which are independent of each other) run sequentially from $0 \to m$ and from $m \to 0$. One can derive an expression for $E_0(T^n|M=m)$ on the basis of (13) and (14);

$$E_0(T^n|M=m) = \sum_{k=0}^{n} \binom{n}{k} E_{x=m}(T^k|M=m) E_{x=m}(T^{n-k}|M=m)$$

$$= \frac{16 n!}{2^n \pi^{2n}} \sum_{k=0}^{n} (2^{2k-1}-1)(2^{2n-2k-1}-1)\zeta(2k)\zeta(2n-2k)m^{2n} \tag{15}$$

where $\zeta(0) \equiv -\frac{1}{2}$ by analytic continuation [15; Sect. 23.2.11]. A much simpler representation, however, may be obtained by employing (4) to derive,

$$\int_0^\infty \Pr(M \leq m, T = t, x) t^n dt = n! m^{2n} \frac{2^{n+1}}{\pi^{2n+1}} \sum_{k=1}^\infty \sin\left(\frac{k\pi x}{m}\right) \frac{1}{k^{2n+1}} \tag{16}$$

$$= xn! m^{2n-1} \frac{2^{n+1}}{\pi^{2n}} \sum_{k=1}^\infty \frac{1}{k^{2n}} + O(x^2).$$

In carrying out this procedure it is permissible to interchange the summation and the integral. Noting once more that $P(m,t,x) \equiv \frac{\partial}{\partial m} \Pr(M \leq m, T = t, x)$ one therefore has,

$$\int_0^\infty t^n P(m,t,x) dt = x(2n-1)n! m^{2n-2} \frac{2^{n+1}}{\pi^{2n}} \sum_{k=1}^\infty \frac{1}{k^{2n}} + O(x^2). \tag{17}$$

The quantity $E_x(T^n | M = m)$ is formally defined by,

$$E_x(T^n | M = m) \equiv \frac{\int_0^\infty t^n P(m,t,x) dt}{P(m,x)} \tag{18}$$

and by combining (9) and (17) we finally have the result we have been seeking;

$$E_0(T^n | M = m) = \frac{(2n-1)2^{n+1} n!}{\pi^{2n}} \zeta(2n) m^{2n}. \tag{19}$$

This agrees with the results in (11) in the appropriate limit $x \to 0$. Comparison of (15) and (19) reveals an interesting identity involving the Riemann zeta function; again, further insights may be found in [10].

We now provide, using the above results, a derivation of $E_0(M^n|T=t)$ which is the counterpart to (19) relevant to Brownian excursions. The derivation is quite different to that based on the results given in [1-3]. The desired quantity is defined by taking the limit $x \to 0$ of the quantity,

$$E_x(M^n|T=t) \equiv \frac{\int_x^\infty m^n P(m,t,x)\,dm}{P(t,x)} \tag{20}$$

where $P(t,x) \equiv \int_x^\infty P(m,t,x)\,dm$ is the probability density of the first passage time. The latter is standard and follows quickly from (1) and (3);

$$\int_0^\infty e^{-pt} P(t,x)\,dt = \int_x^\infty \left( \int_0^\infty e^{-pt} P(m,t,x)\,dt \right) dm = \int_x^\infty \frac{\partial Q(m,p,x)}{\partial m}\,dm$$

$$= \lim_{m \to \infty} Q(m,p,x) = e^{-\sqrt{2p}\,x} \tag{21}$$

whose inversion yields (see e.g. [12]),

$$P(t,x) = \frac{x}{\sqrt{2\pi}} \frac{1}{t^{3/2}} e^{-x^2/2t}. \tag{22}$$

Next, from (4) we have that,

$$P(m,t,x) = \pi^2 x \frac{\partial}{\partial m}\left[ \frac{1}{m^3} \sum_{k=1}^\infty k^2 e^{-\frac{k^2\pi^2}{2m^2}t} \right] + O(x^2). \tag{23}$$

This form is not suitable to evaluate the integral in the numerator of (20) as it stands. This is because interchanging the summation and the integral leads to a divergent result and is therefore not permitted. Instead, one can make use of the Jacobi theta function identity (Poisson summation formula),

$$\sum_{k=-\infty}^{\infty} e^{-\frac{k^2\pi^2}{2m^2}t} = \left(\frac{2m^2}{\pi t}\right)^{1/2} \sum_{k=-\infty}^{\infty} e^{-\frac{2k^2m^2}{t}} \quad (24)$$

which, after differentiating both sides with respect to $m$ and rearranging, yields,

$$\frac{1}{m^3}\sum_{k=1}^{\infty} k^2 e^{-\frac{k^2\pi^2}{2m^2}t} = \frac{1}{2^{1/2}\pi^{5/2}t^{3/2}}\left[1 + 2\sum_{k=1}^{\infty} e^{-\frac{2k^2m^2}{t}} - \frac{8m^2}{t}\sum_{k=1}^{\infty} k^2 e^{-\frac{2k^2m^2}{t}}\right]. \quad (25)$$

Substituting for the corresponding term in (23) and then evaluating (20), recognising that for $n > 1$ it is now possible to interchange the summation and the integral without generating divergent results, gives in the limit $x \to 0$,

$$E_0(M^n | T = t) = \frac{32}{t^2}\sum_{k=1}^{\infty} k^4 I_{n+3}(k,t) - \frac{24}{t}\sum_{k=1}^{\infty} k^2 I_{n+1}(k,t);$$

$$I_{n+\alpha}(k,t) \equiv \int_0^{\infty} m^{n+\alpha} e^{-\frac{2k^2m^2}{t}} dm \quad (26)$$

$$= \frac{1}{2}\left(\frac{t}{2k^2}\right)^{(n+\alpha+1)/2} \Gamma\left(\frac{n+\alpha+1}{2}\right)$$

which reduces to the final result [4, 10],

$$E_0(M^n|T=t) = \frac{n(n-1)\Gamma(\frac{n}{2})}{2^{n/2}}\zeta(n)t^{n/2}. \tag{27}$$

The first moment $E_0(M|T=t) = \sqrt{\pi t/2}$ may be obtained from (27) by noting the limiting behaviour $\lim_{n\to 1^+}(n-1)\zeta(n) = 1$ [15; Sect. 23.2.5].

## 3. Motion conditioned by area

We now turn to the more challenging problem of Brownian motion conditioned by area. The conditional expectation $E_x(T^n|A=a)$ can be derived from the joint probability density $P(a,t,x)$ associated with the distribution $\Pr(A \leq a, T \leq t, x)$. This time we study the double Laplace transform of $P(a,t,x)$,

$$R(s,p,x) = \int_0^\infty \int_0^\infty e^{-sa}e^{-pt}P(a,t,x)\,da\,dt \tag{28}$$

which satisfies the backward Fokker-Plank equation (see Appendix);

$$\left[\frac{1}{2}\frac{\partial^2}{\partial x^2} - (sx+p)\right]R(s,p,x) = 0. \tag{29}$$

The relevant boundary conditions are $R(s,p,x=0) = 1$ and $R(s,p,x\to\infty) = 0$, whereupon the solution of (29) may be expressed in terms of the Airy function $\text{Ai}(z)$,

$$R(s, p, x) = \frac{\text{Ai}\left(\frac{2^{1/3}p + 2^{1/3}sx}{s^{2/3}}\right)}{\text{Ai}\left(\frac{2^{1/3}p}{s^{2/3}}\right)}. \tag{30}$$

The quantity $E_x(e^{-pT}|A=a)$ is formally defined by,

$$E_x(e^{-pT}|A=a) \equiv \frac{\int_0^\infty e^{-pt} P(a,t,x)\,dt}{P(a,x)} \tag{31}$$

where $P(a,x) \equiv \int_0^\infty P(a,t,x)\,dt$ is the area probability density. Since from (28) we have that $R(s, p=0, x) \equiv E_x(e^{-sA})$, this may be obtained directly from (30) after setting $p=0$ and inverting with respect to $s$ [14];

$$P(a,x) = \frac{2^{1/3}}{3^{2/3}\Gamma(\frac{1}{3})} \frac{x}{a^{4/3}} e^{-2x^3/9a}. \tag{32}$$

Written another way, it follows from (28) and (31) that,

$$\int_0^\infty e^{-sa} E_x(e^{-pT}|A=a) P(a,x)\,da = R(s,p,x). \tag{33}$$

To determine $E_x(e^{-pT}|A=a)$ it is necessary to carry out the inverse Laplace transform of (33) with respect to $s$. This is technically non-trivial, but following the approach in [16] it can be accomplished (in the sense of reducing the final form to a real integral) by deforming the conventional Bromwich contour to a Hankel-type contour around the branch cut $(-\infty, 0]$. By considering the contributions above and below the branch cut one finds that

$$E_x(e^{-pT}|A=a)P(a,x) = \frac{1}{2\pi i}\int_0^\infty \left[\frac{\text{Ai}(2^{1/3}e^{i2\pi/3}r^{-2/3}(p+e^{-i\pi}xr))}{\text{Ai}(2^{1/3}pe^{i2\pi/3}r^{-2/3})}\right.$$

$$\left. - \frac{\text{Ai}(2^{1/3}e^{-i2\pi/3}r^{-2/3}(p+e^{i\pi}xr))}{\text{Ai}(2^{1/3}pe^{-i2\pi/3}r^{-2/3})}\right]e^{-ar}dr. \tag{34}$$

Using the identity [15; Sect. 10.4.9]

$$\text{Ai}(ze^{\pm 2\pi i/3}) = \frac{1}{2}e^{\pm \pi i/3}[\text{Ai}(z) \mp i\text{Bi}(z)] \tag{35}$$

and using (32) one can recast (34) into the form

$$E_x(e^{-pT}|A=a) = \frac{3^{2/3}\Gamma(\tfrac{1}{3})a^{4/3}e^{2x^3/9a}}{\pi 2^{1/3}x}\int_0^\infty \frac{f_x(r,p)e^{-ar}}{\text{Ai}^2(2^{1/3}pr^{-2/3})+\text{Bi}^2(2^{1/3}pr^{-2/3})}dr$$

$$\tag{36}$$

$$f_x(r,p) = \text{Ai}(2^{1/3}pr^{-2/3} - x2^{1/3}r^{1/3})\text{Bi}(2^{1/3}pr^{-2/3})$$

$$- \text{Bi}(2^{1/3}pr^{-2/3} - x2^{1/3}r^{1/3})\text{Ai}(2^{1/3}pr^{-2/3}).$$

It is clear from this that there is no simple expression for $E_x(T^n|A=a)$ for a general value of $x$. However, noting the fact that $\text{Ai}(z)\text{Bi}'(z) - \text{Ai}(z)'\text{Bi}(z) = 1/\pi$ [15; Sect. 10.4.10], one can deduce in the limit $x \to 0$ that $f_x(r,p) = x2^{1/3}r^{1/3}/\pi + O(x^2)$ and therefore,

$$E_0(e^{-pT}|A=a) = \frac{3^{2/3}\Gamma(\tfrac{1}{3})}{\pi^2}\int_0^\infty \frac{y^{1/3}e^{-y}dy}{\operatorname{Ai}^2(2^{1/3}pa^{2/3}y^{-2/3})+\operatorname{Bi}^2(2^{1/3}pa^{2/3}y^{-2/3})}. \qquad (37)$$

From this one can derive $E_0(T|A=a)$ by differentiating once with respect to $p$, setting $p=0$ and then evaluating the integral (the order of these operations can be justified in this special case). This gives after appropriate simplification a key result,

$$E_0(T|A=a) = \frac{4\pi^2 6^{1/3}}{\Gamma(\tfrac{1}{3})^4}a^{2/3} = 1.3928\ldots a^{2/3}. \qquad (38)$$

To physically motivate this let us consider the random acceleration process, where now $X(\tau)$ denotes particle velocity (rather than spatial position) and $A$ is the distance travelled up to the time $T$ when the velocity first reaches zero. If one observes a large number of particles (which all start at the origin with velocity $x$) and focuses on those that first stop moving at a given distance $A=a$ from the origin, then (38) tells us the average time they have taken to arrive (assuming $x$ is relatively small, i.e. $x \ll a^{1/3}$). Related ideas but in a much more general setting are discussed in [17, 18, 19]. In figure 3 we show (on a logarithmic scale) the results of simulations and the expected value of $T$ for a given value of $A=a$ based on (38), focussing on values large enough $a \gg x^3$ to ensure that the non-zero initial condition ($x=1$, as for figure 1 and figure 2) plays no significant role. The agreement is again excellent. Incidentally, from (36) and (38) one can deduce that the expression for $E_x(T|A=a)$ given previously in [20] is incorrect. The error may be traced to a mistake in the manipulation of an equation which is analogous to (33).

Unfortunately, one cannot easily calculate $E_0(T^n|A=a)$ for general $n$ from (37) by differentiating $n$ times with respect to $p$. This is because for $n>1$ the act of differentiating and setting $p=0$ before evaluating the integral leads to a divergent result and is therefore not permitted, and evaluating the integral first is highly non-trivial. The easiest way to proceed is to go back to (33) and consider the limit $x \to 0$ from the outset. Thus we have using (30),

$$\int_0^\infty e^{-sa} E_x(e^{-pT}|A=a) P(a,x) da = 1 + 2^{1/3} s^{1/3} f\left(\frac{2^{1/3} p}{s^{2/3}}\right) x + O(x^2) \qquad (39)$$

where $f(z) \equiv \text{Ai}'(z)/\text{Ai}(z)$. This function has a well-defined Taylor series about the origin;

$$f(z) = \sum_{k=0}^\infty (-1)^k C_k z^k; \qquad C_0 \equiv \frac{\text{Ai}'(0)}{\text{Ai}(0)} = -\frac{3^{1/3} \Gamma(\tfrac{2}{3})}{\Gamma(\tfrac{1}{3})} \qquad (40)$$

where, since $f(z)$ satisfies the differential equation $f'(z) = z - f(z)^2$, one has $C_1 = C_0^2$ and $C_2 = C_0^3 + \tfrac{1}{2}$, whilst for $n>2$ one can determine the coefficients $C_n$ recursively;

$$C_n = \frac{1}{n} \sum_{j=1}^n C_{j-1} C_{n-j}. \qquad (41)$$

Thus one can expand the $O(x)$ term on right hand side of (39) as a power series in $p$, rendering the subsequent step of comparing coefficients straightforward;

$$\int_0^\infty e^{-sa} E_x(T^n | A = a) P(a, x) da = \frac{2^{(n+1)/3} n!}{s^{(2n-1)/3}} C_n x + O(x^2). \tag{42}$$

The inversion with respect to $s$ is now elementary to $O(x)$ and, after using (32) and taking the limit $x \to 0$, we have the main result of this Section;

$$E_0(T^n | A = a) = \frac{3^{2/3} \Gamma(\frac{1}{3}) 2^{n/3} n!}{\Gamma(\frac{2n-1}{3})} C_n a^{2n/3}. \tag{43}$$

After some modest algebraic manipulation it is easy to show that this agrees with (38) when $n = 1$. The result is complete apart from the fact that there is no known closed-form solution for the coefficient $C_n$. In figure 4, we complement figure 3 and compare (on a logarithmic scale) the standard deviation $\sigma_a(T) \equiv \sqrt{\text{Var}(T)}_{A=a}$ based on (43) against numerical averages derived from simulations, with good agreement.

The moments (43) are the counterpart to the result for excursions $E_0(A^n | T = t)$. To make the connection one can mirror the procedure used to derive (33) to obtain,

$$\int_0^\infty e^{-pt} E_x(e^{-sA} | T = t) P(t, x) dt = R(s, p, x) \tag{44}$$

where $P(t,x)$ is the first passage time probability density given by (22). We again expand the right hand side of (44) to $O(x)$ as was done in (39), but this time we note that $f(z)$ has a well-defined asymptotic expansion as $z \to \infty$ [16];

$$f(z) \sim 2z^{1/2} \sum_{n=0}^{\infty} (-1)^n K_n z^{-3n/2} \qquad (45)$$

where $K_0 = -\frac{1}{2}$, $K_1 = \frac{1}{8}$ whilst for $n > 1$ one has the quadratic recursion [9],

$$K_n = \left(\frac{3n-4}{4}\right) K_{n-1} + \sum_{j=1}^{n-1} K_j K_{n-j}. \qquad (46)$$

An explicit expression for $K_n$ is given in [16] in terms of an integral involving Airy functions. It follows that the $O(x)$ term can be expanded as a power series in $s$, whereupon after comparing coefficients and carrying out the Laplace inversion with respect to $p$ one has [7, 9],

$$E_0(A^n | T = t) = \frac{4\sqrt{\pi} n!}{\Gamma(\frac{3n-1}{2}) 2^{n/2}} K_n t^{3n/2}. \qquad (47)$$

For ease of reference and for comparison with the counterpart result (38), the first moment is given by $E_0(A | T = t) = \sqrt{\pi/8}\, t^{3/2}$.

## 4. Discussion

In relation to the first passage process $X(\tau) = x + W(\tau)$, the unconditional moments of $T$, $M$ and $A$ are all infinite, as may be inferred from the corresponding densities (9), (22) and (32). Thus to characterise the process it is helpful to examine such quantities conditionally. This we have done by calculating the moments of $T$ in the limit $x \to 0$ conditioned on maximum value $E_0(T^n|M=m)$ and area $E_0(T^n|A=a)$. These results are directly useful in that, if one is given either the maximum value of the process or the area defined by the process, one can estimate the duration of the motion. As an illustration, in relation to the problem of linear probing with random hashing given in the Introduction, one can estimate using (38) the size of the storage table given the total cost function, and this is, in principle, a relevant consideration from an algorithmic point of view [21]. Applications in biology and ecology relating to random exploration within a constrained environment are also suggestive, but the extension of the basic ideas to higher spatial dimensions is not straightforward.

The spread of the data around the mean value as depicted in figures 1-4 is relevant and the following results, deduced from (19) and (43), are of interest;

$$\left.\frac{\sigma(T)}{E(T)}\right|_{M=m} = \frac{1}{\sqrt{5}} = 0.4472\ldots \qquad \left.\frac{\sigma(T)}{E(T)}\right|_{A=a} = 0.1628\ldots \qquad (48)$$

These hold exactly for $x = 0$ and asymptotically as $m \to \infty$ or $a \to \infty$ for $x > 0$. It is clear that the first passage time is more tightly constrained by fixing the area than the maximum value, which is intuitively what one would expect.

As a final observation, by integrating (14) with respect to $p$ over the open interval $[0, \infty)$ one can derive,

$$E_0(T^{-1}|M = m) = \int_0^\infty \frac{2pm^2}{\sinh^2(\sqrt{2}pm)} dp = \frac{3\zeta(3)}{2} m^{-2}. \tag{49}$$

This result, which may also be deduced from (19) by analytic continuation, reveals an unexpected physical manifestation of Apery's constant $\zeta(3)$. One can similarly exploit (37) in the same way to derive,

$$E_0(T^{-1}|A = a) = \frac{\Gamma(\frac{1}{3})}{2^{4/3} 3^{1/3}} a^{-2/3} \tag{50}$$

where, after interchanging the order of integration, use has been made of the standard integral $\int_0^\infty [\mathrm{Ai}^2(z) + \mathrm{Bi}^2(z)]^{-1} dz = \pi^2/6$ [16]. Both (49) and (50) agree well with the results of simulations.

**Acknowledgement**

MJK acknowledges useful discussions with Andrew Pye in respect of the simulations. SNM acknowledges support by ANR grant 2011-BS04-013-01 WALKMAT and in part by the Indo-French Centre for the Promotion of Advanced Research under Project 4604-3.

**Appendix. Derivation of the backward Fokker-Planck equations**

Define the general functional expectation

$$F(p,x) = \left\langle \exp\left(-p\int_0^T U(X(\tau))d\tau\right)\right\rangle \tag{A1}$$

where $\langle...\rangle$ denotes the average over all allowed paths of the process $X(\tau)$, starting at $x$, until their first-passage time and $U(z)$ is an arbitrary positive function. To evaluate $F$, we split a typical path over the interval $[0,T]$ into two parts: a left interval $[0,\Delta\tau]$ where the paths moves from $x$ to $x+\Delta X$ in a small time interval $\Delta\tau$ and a right interval $[\Delta\tau,T]$ during which the path starts at $x+\Delta X$ at time $\Delta\tau$ and crosses the origin for the first time in the interval $[T,T+\Delta T]$. The integral $\int_0^T U(X(\tau))d\tau$ can then also be split into two parts: $\int_0^T = \int_0^{\Delta\tau} + \int_{\Delta\tau}^T$. Since the initial condition is $x$ we have $\int_0^{\Delta\tau} U(X(\tau))d\tau = U(x)\Delta\tau + O(\Delta\tau^2)$ as $\Delta\tau \to 0$. The Markov nature of the process then means that (A1) can be written as,

$$F(p,x) = \left\langle \exp(-pU(x)\Delta\tau)F(p,x+\Delta X)\right\rangle_{\Delta X} + O(\Delta\tau^2) \tag{A2}$$

where the average is now taken over the incremental displacement $\Delta X$. Expanding $F(p,x+\Delta X)$ to $O(\Delta X^2)$, and noting that for the Wiener process $\langle \Delta X\rangle = 0$ and $\langle \Delta X^2\rangle = \Delta\tau$, one obtains from (A2),

$$\left[\frac{1}{2}\frac{\partial^2 F}{\partial x^2} - pU(x)F\right]\Delta\tau + O(\Delta\tau^2) = 0. \tag{A3}$$

Dividing through by $\Delta\tau$ and taking the limit $\Delta\tau \to 0$ one therefore sees that $F$ obeys the backward Fokker-Planck equation,

$$\frac{1}{2}\frac{\partial^2 F}{\partial x^2} - pU(x)F = 0. \tag{A4}$$

In the case where we are interested in the maximum of the process, we take $U(z) = 1$ and note that the initial variable $x \in [0, m]$. It follows from the definition (A1) that $F = \langle e^{-pT}\rangle_{M \leq m} \equiv Q(m, p, x)$ and (A4) reduces to (2) in the main text. Regarding the boundary conditions: (i) as the initial position $x \to 0$ so the first passage time $T \to 0$ almost surely, hence from (A1) we have $Q(m, p, x=0) = 1$; (ii) as $x \to m$ the paths for which $M \leq m$ have vanishing measure and so $Q(m, p, x=m) = 0$. In the case where we are interested in the area swept out by the process, we take $U(z) = 1 + (s/p)z$ and note that the initial variable $x \in [0, \infty)$. It follows that $F = \langle e^{-sA-pT}\rangle \equiv R(s, p, x)$ and (A4) reduces to (29) in the main text. Regarding the boundary conditions: (i) as before, as $x \to 0$ so $T \to 0$ almost surely and hence $R(m, p, x=0) = 1$; (ii) as $x \to \infty$ so $T \to \infty$ and consequently the integral $\int_0^T (1 + (s/p)X(\tau))d\tau \to \infty$, which implies $R(m, p, x \to \infty) = 0$. Further insights into such methods and the Feynman-Kac formalism in general may be found in [8, 9].

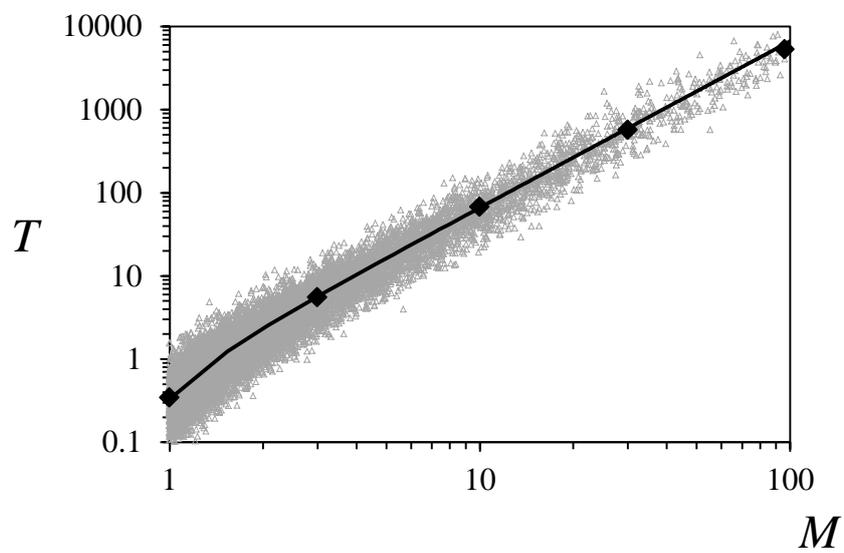

Figure. 1. Comparison between the expected value of *T* for a given value of *M* based on theory (solid line) and numerical averages (solid diamonds) derived from simulations (open triangles).

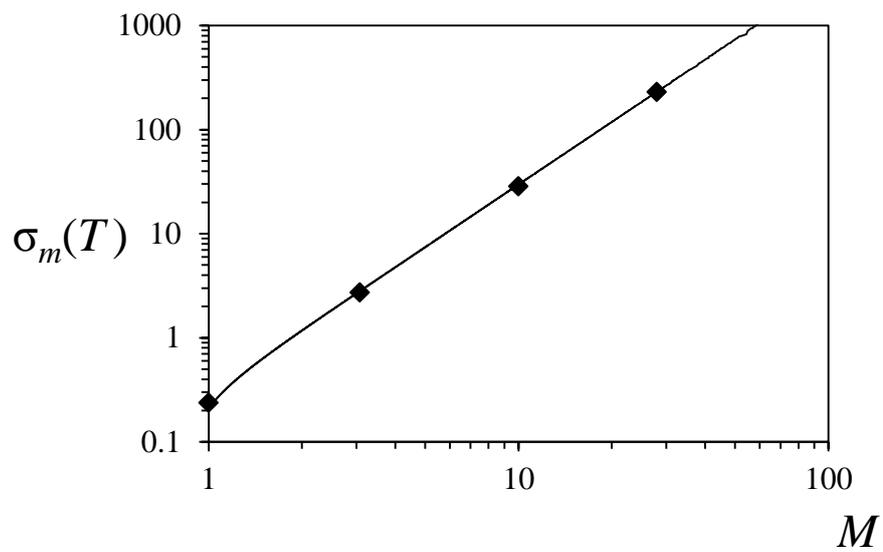

Figure. 2. Comparison between the standard deviation of *T* for a given value of *M* based on theory (solid line) and numerical averages (solid diamonds) derived from simulations.

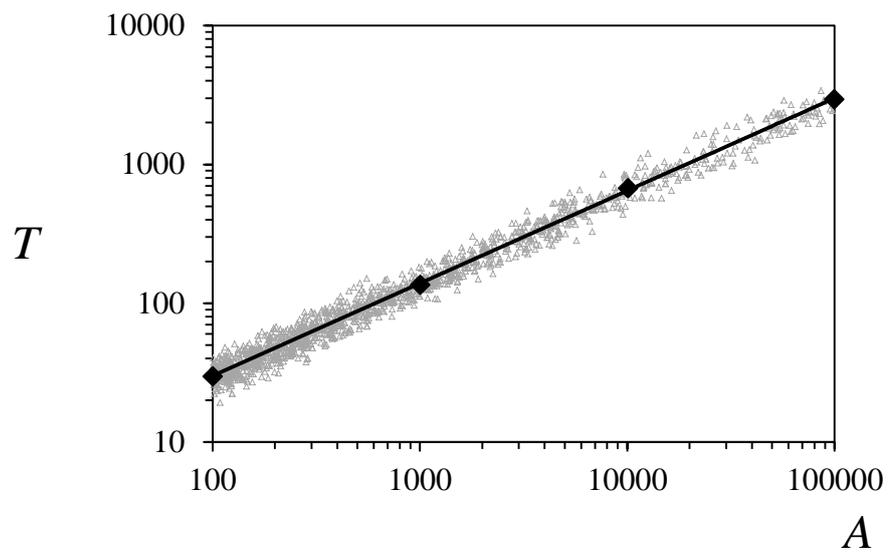

Figure. 3. Comparison between the expected value of *T* for a given value of *A* based on theory (solid line) and numerical averages (solid diamonds) derived from simulations (open triangles).

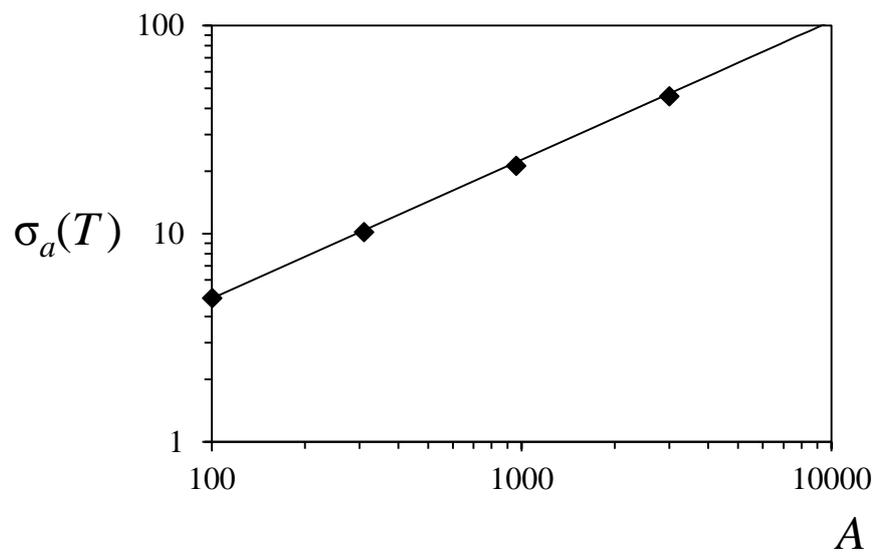

Figure. 4. Comparison between the standard deviation of *T* for a given value of *A* based on theory (solid line) and numerical averages (solid diamonds) derived from simulations.